\documentclass[reprint,amsmath,aip,pop]{revtex4-2}
\pdfoutput=1
\usepackage[left=1in,top=1in,right=1in,bottom=1in]{geometry}
\usepackage{graphicx}
\usepackage{color}
\usepackage{hyperref}
\usepackage{natbib}
\usepackage[english]{babel}
\usepackage[usenames,dvipsnames,svgnames]{xcolor}
\usepackage{float,tabulary}
\usepackage{booktabs, bm}
\usepackage{subcaption}
\usepackage{multirow}
\graphicspath{{./figures/}}

\newcommand{\neped}{n_{e,ped}}
\newcommand{\betan}{\beta_N}

\newcommand{\hcd}{H{\&}CD}
\newcommand{\tabletab}[1]{\quad{\footnotesize #1}}
\newcommand{\nt}{negative-$\delta$}
\newcommand{\pt}{positive-$\delta$}

\begin{document}
\title{Flexible, integrated modeling of tokamak stability, transport, equilibrium, and pedestal physics}

\author{B.C. Lyons}
\affiliation{General Atomics, San Diego, California 92121, USA}
\author{J. McClenaghan}
\affiliation{General Atomics, San Diego, California 92121, USA}
\author{T. Slendebroek}
\affiliation{Oak Ridge Institute for Science and Education, Oak Ridge, Tennessee 37831-0117, USA}
\affiliation{General Atomics, San Diego, California 92121, USA}
\author{O. Meneghini}
\affiliation{General Atomics, San Diego, California 92121, USA}
\author{T.F. Neiser}
\affiliation{General Atomics, San Diego, California 92121, USA}
\author{S.P. Smith}
\affiliation{General Atomics, San Diego, California 92121, USA}
\author{D.B. Weisberg}
\affiliation{General Atomics, San Diego, California 92121, USA}
\author{E.A. Belli}
\affiliation{General Atomics, San Diego, California 92121, USA}
\author{J. Candy}
\affiliation{General Atomics, San Diego, California 92121, USA}
\author{J.M. Hanson}
\affiliation{Columbia University, New York, New York 10027, USA}
\author{L.L. Lao}
\affiliation{General Atomics, San Diego, California 92121, USA}
\author{N.C. Logan}
\affiliation{Lawrence Livermore National Laboratory, Livermore, California 94550, USA}
\author{S. Saarelma}
\affiliation{UK Atomic Energy Authority, Culham Science Centre, Abingdon OX14 3DB, United Kingdom}
\author{O. Sauter}
\affiliation{\'{E}cole Polytechnique F\'{e}d\'{e}rale de Lausanne (EPFL), Swiss Plasma Center (SPC), CH-1015 Lausanne, Switzerland}
\author{P.B. Snyder}
\affiliation{Oak Ridge National Laboratory, Oak Ridge, Tennessee 37831, USA}
\author{G.M. Staebler}
\affiliation{General Atomics, San Diego, California 92121, USA}
\author{K.E. Thome}
\affiliation{General Atomics, San Diego, California 92121, USA}
\author{A.D. Turnbull}
\affiliation{General Atomics, San Diego, California 92121, USA}

\begin{abstract}
The STEP (Stability, Transport, Equilibrium, and Pedestal) integrated-modeling tool has been developed in OMFIT to predict stable, tokamak equilibria self-consistently with core-transport and pedestal calculations. STEP couples theory-based codes to integrate a variety of physics, including MHD stability, transport, equilibrium, pedestal formation, and current-drive, heating, and fueling. The input/output of each code is interfaced with a centralized ITER-IMAS data structure, allowing codes to be run in any order and enabling open-loop, feedback, and optimization workflows. This paradigm simplifies the integration of new codes, making STEP highly extensible. STEP has been verified against a published benchmark of six different integrated models. Core-pedestal calculations with STEP have been successfully validated against individual DIII-D H-mode discharges and across more than 500 discharges of the $H_{98,y2}$ database, with a mean error in confinement time from experiment less than 19\%. STEP has also reproduced results in less conventional DIII-D scenarios, including negative-central-shear and negative-triangularity plasmas. Predictive STEP modeling has been used to assess performance in several tokamak reactors. Simulations of a high-field, large-aspect-ratio reactor show significantly lower fusion power than predicted by a zero-dimensional study, demonstrating the limitations of scaling-law extrapolations. STEP predictions have found promising EXCITE scenarios, including a high-pressure, 80\%-bootstrap-fraction plasma. ITER modeling with STEP has shown that pellet fueling enhances fusion gain in both the baseline and advanced-inductive scenarios. Finally, STEP predictions for the SPARC baseline scenario are in good agreement with published results from the physics basis.
\end{abstract}

\maketitle

\section{Introduction}
\label{sec:intro}

World-wide fusion research is moving towards a new paradigm with a significantly increased emphasis on energy production. ITER\cite{ikeda:2007} construction is more than 77\% complete to first plasma\cite{iter:2022}. It is expected to begin pre-fusion-power operation in the late 2020s and then to begin deuterium-tritium (DT) fusion experiments in the mid-2030s, demonstrating a fusion gain ($Q$) of at least ten. The SPARC tokamak\cite{greenwald:2020}, being built by the U.S. private sector, is targeting D-T operation to demonstrate net-energy production from fusion in the mid 2020s. In addition, recent U.S. strategic-planning activities\cite{cpp:2020, fesac:2020} and National Academy of Sciences studies\cite{nasem:2019, nasem:2021} have converged on a fusion pilot plant mission for the United States, with the goal of net electricity production in the mid-to-late 2030s.

The high capital cost of fusion devices prevents design of reactors through iterative building and testing. Once a reactor is constructed, it must meet the performance milestones it set out to achieve. Thus, the success of future tokamak reactors will rely on modeling to predict configurations that have high performance and are stable against disruptive magnetohydrodynamic (MHD) instabilities that terminate plasma confinement and can damage the reactor. Typically, such modeling is done at a variety of fidelities, beginning with zero-dimensional (0D) systems codes and/or scaling laws to map out a high-dimensional parameter space of potential design points and culminating with high-fidelity, first-principle simulations of individual physical phenomena (e.g., gyrokinetic turbulent-transport or 3D, nonlinear MHD). In between these two extremes are reduced models that are significantly faster than first-principle codes but have greater fidelity than 0D models.

Tokamaks have a wide-variety of physical phenomena that must harmonize for successful operation. MHD accounts for both the equilibrium state of the plasma and the means by which that equilibrium can go partially or completely unstable. Drift-kinetics and turbulence combine to transport particles, momentum, and heat through the core plasma. Drift-kinetics, radio-frequency wave injection, neutral-beam injection, and resistive diffusion account for the current profile within the plasma. Fast particles provide a source of heating and potential instability. Heat and particles are exhausted from the core into a scrape-off layer and terminating in a divertor. In those regions alone, one must account for turbulence, particle drifts, neutral dynamics, radiation, and plasma-material interaction. The list above is certainly incomplete and consists only of plasma-physics phenomena. A true whole-device model would also account of interactions with the tokamak hardware, including material science and neutronics. Model integration is thus essential to predict tokamak behavior accurately.

We have developed a new integrated modeling tool called STEP in the OMFIT integrated-modeling framework\cite{meneghini:2013,meneghini:2015}. STEP was written to be flexible, allowing users to select which codes to run and in which order to run them, as well as extensible, allowing for the addition of new codes and models with minimal effort. These features allow STEP to create a variety of workflows with different levels of fidelity. The existing, broad user-base of OMFIT combined with these user-friendly features will allow STEP to be a valuable integrated-modeling tool for the fusion community. In addition, while STEP is currently capable of accurately modeling steady-state, core plasma behavior, the above features will allow it to incorporate any and all phenomena needed for a whole-device model.

Here we provide an overview of STEP's capabilities and some initial applications. In Section \ref{sec:step}, we describe STEP in detail, focusing on its use of centralized data exchange and model modularity. We also describe some of the standard workflows that have been built and are routinely utilized in STEP. In Section \ref{sec:v&v}, we show that STEP has been verified against predictive simulations by other integrated-modeling workflows and experimentally validated against both standard H-mode and less conventional tokamak scenarios. In Section \ref{sec:reactor}, we demonstrate STEP's predictive capabilities, performing simulations for ITER, SPARC, and EXCITE (a next-step, high-performance, research tokamak proposed by the recent strategic-planning activities). Section \ref{sec:concl} summarizes this research and provides an outlook on future work.

\section{STEP Integrated-Modeling Tool}
\label{sec:step}

\begin{figure}
\centering
\includegraphics[width=0.95\columnwidth]{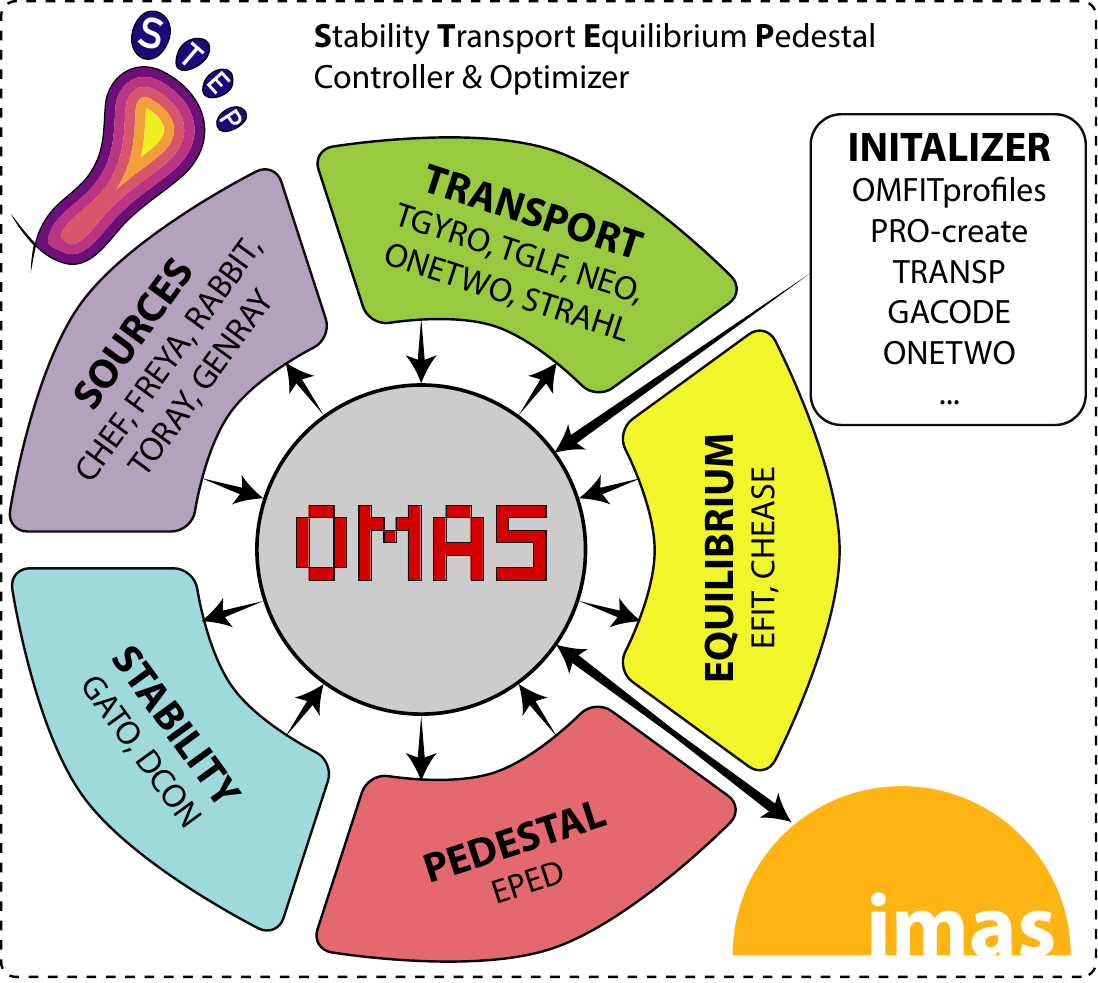}
\caption{A schematic of STEP. Various physics codes and models can be run in any desired order by interfacing with an OMAS-based centralized data structure. STEP can be initialized from a variety of existing codes and data formats and its results can be directly written to ITER's IMAS.}
\label{fig:STEP}
\end{figure}

STEP couples together \underline{S}tability, source, \underline{T}ransport, \underline{E}quilibrium, and \underline{P}edestal codes to predict self-consistent, stationary tokamak plasma states. It has been implemented in Python through OMFIT, allowing access to a wide-variety of physics modules and codes, as well as an established, broad user-base. Rather than linking specific codes in static, predefined workflows, STEP provides a platform by which users can develop their own workflows using any number of physics codes run in any desired order. This is accomplished by wrapping each code into individual ``steps'' that read from and write to a centralized data structure. Thus, the flexibility of STEP is enabled by two core concepts: centralized data exchange and model modularity. A brief description of STEP was included in Ref. \onlinecite{meneghini:2020}, but we provide more details here.

\subsection{Centralized data exchange}
\label{sec:omas}

In general, every physics code or model uses its own data formats for input and output; linking them together requires translation routines so the codes can communicate. Fixed workflows with predefined codes and execution order need only to create translators between adjacent codes, but new translators must be created each time the workflow is changed. For such a formulation, coupling $N$ codes in arbitrary order would require $N(N-1)$ translators and adding an additional code would require $2N$ translators. STEP avoids these complications by having every code read its inputs from and write its outputs to a centralized data structure, necessitating only $2N$ translators total. This approach has been pursued over the years by different projects within the fusion community, including TRANSP/IPS plasma state\cite{elwasif:2010}, ITM consistent physical objects\cite{falchetto:2014}, GACODE input file \cite{gacode:2022}, and TASK Burning Plasma Simulation Initiative\cite{fukuyama:2011}. Each of these projects' data structures, however, are incompatible with one another and are not used outside of their original project.

STEP's centralized data structure is based on the ITER Integrated Modelling \& Analysis Suite (IMAS) data schema\cite{imbeaux:2015}, which provides a standardized format for both experimental and simulated data and has emerged as a de facto standard for fusion data in recent years. IMAS data is organized hierarchically into individual Interface Data Structures (IDSs), 68 at present, that cover different aspects of the tokamak plasma state (e.g., equilibrium, core and edge profiles, core and edge sources). STEP uses the Ordered Multidimensional Array Structures\cite{meneghini:2020} (OMAS) to interface with IMAS. Written in Python, OMAS enforces compatibility with the IMAS data schema, while providing automated features like interpolation, coordinate-convention (COCOS\cite{sauter:2013}) conversion, and self-consistent derivation of physical quantities (e.g., pressures computed from provided densities and temperatures). Every code step in STEP, therefore, reads from and writes to an OMAS data structure (ODS). While some of this is done within STEP itself, the process is largely automated within OMFIT, which can translate between an ODS and a wide variety of typical file formats, (e.g., EFIT's gEQDSK and GACODE's input.gacode). STEP stores the ODS after every step taken, giving a complete history of every run through every code used.

\subsection{Modularity and physics ``steps''}
\label{sec:modularity}

STEP uses OMFIT modules to access various physics codes and models. The modules allow for easy access to and editing of input files, execution of codes on both local computers and remote servers, and collection of output results. Each module is further wrapped into its own ``step,'' which automates the running of the module. First, data is read from the previous step's ODS and the inputs needed for the present step are setup. Some of this is handled by the automated file translation described in Sec. \ref{sec:omas}, but STEP also initializes input files and codes with a robust set of default settings. After this, each code is executed in an automated way using a combination of routines created for the specific OMFIT module as well as custom coding for STEP-specific behavior. Finally, outputs of the step are translated to an ODS, updating existing data in the ODS as appropriate. These three phases of a step (i.e., setup, run, and save) can be executed automatically using all default parameters or one at a time. The latter allows expert users to change the inputs to a step between setting up and running or to check the outputs between running and saving, providing an opportunity for more detailed control and customization.

\begin{table*}
\centering
\begin{tabular}{llcc} 
 \hline
 \hline
Physics step & Brief Description & Inputs & Outputs  \\ 
\hline
DCON\cite{glasser:1997} & ideal MHD stability & eq. & perturbed energy ($\delta W$) \\
GATO\cite{bernard:1981} & ideal MHD stability & eq. & growth rates \\
ONETWO\cite{pfeiffer:1980, stjohn:1994} & 1.5D transport  & eq., profiles, [sources] & current profile, sources \\
\tabletab{NUBEAM\cite{goldston:1981,pankin:2004}} & \tabletab{ neutral beams \hcd{}} & & \\
\tabletab{FREYA\cite{stjohn:1994}} & \tabletab{ neutral beams \hcd{}} & & \\
\tabletab{TORAY\cite{kritz:1982,matsuda:1989}} & \tabletab{ electron-cyclotron \hcd{}} & & \\
CHEF & \hcd{}, fueling  & eq., profiles & sources \\
\tabletab{FREYA\cite{stjohn:1994}} & \tabletab{ neutral beams \hcd{}} & & \\
\tabletab{RABBIT\cite{weiland:2018,weiland:2019}} & \tabletab{ neutral beams \hcd{}} & & \\
\tabletab{TORAY} & \tabletab{ electron-cyclotron \hcd{}} & & \\
\tabletab{GENRAY\cite{smirnov:1995}} & \tabletab{ radio-frequency \hcd{}} & & \\
\tabletab{PAM\cite{mcclenaghan:2023}} & \tabletab{ pellet fueling} & & \\

EPED\cite{snyder:2009,snyder:2011}(-NN) & pedestal stability & eq., profiles & pedestal pressure structure \\

NEO\cite{belli:2008, belli:2012} & drift-kinetic solver & eq., profiles & bootstrap current \\
TGYRO\cite{candy:2009} & steady-state transport & profiles, sources & profiles \\
\tabletab{TGLF\cite{staebler:2007}(-NN\cite{meneghini:2017})} & \tabletab{ turbulent transport} & & \\
\tabletab{GLF23\cite{waltz:1997}} & \tabletab{ turbulent transport} & & \\
\tabletab{EPED-NN\cite{meneghini:2017}} & \tabletab{pedestal structure} & & \\
\tabletab{NEO} & \tabletab{neoclassical transport} & & \\
STRAHL\cite{dux:2004} &  impurity transport & eq., profiles & impurity transport \& profiles \\

EFIT\cite{lao:1985, lao:1990, lao:2005} & free-boundary eq. & eq., profiles, current & eq. \\
CHEASE\cite{lutjens:1996} & fixed-boundary eq. & eq., profiles, current & eq. \\

\hline 

 \end{tabular}
\caption{Physics steps (codes and models) currently available within STEP. ``eq.'' refers to Grad-Shafranov equilibrium, ``profiles'' refer to kinetic profiles (like density and temperature), and ``\hcd{}'' refers to heating and current drive. Smaller, indented lines denote that the code is accessible through the above, non-indented step. [$\cdots$] denotes optional input/output. ``-NN'' denotes that a neural-network version of the code is available.} 
 \label{table:steps}
 \end{table*}

Table \ref{table:steps} provides a comprehensive list of the many well-known, state-of-the-art physics codes currently available in STEP, along with citations, a brief description, and the required, physical inputs and outputs for each step. The modular nature described, however, makes STEP readily extensible. To add a new step, one only needs to create an appropriate OMFIT module (if it does not already exist), define translation of the inputs from and outputs to the ODS format, and write a short script to do the automated setup, execution, and saving. We anticipate additional physics steps to be made available regularly. Information on available steps and how to create new steps can be found on the OMFIT website\cite{omfit:2022}.

\subsection{Standard workflows}
\label{sec:workflows}

While STEP users can create any workflow desired, or execute one step at a time for custom, on-the-fly analysis, several default workflows have been created. The primary standard workflow runs ONETWO for sources (including neutral beam injection [NBI] and radio-frequency [RF] heating \& current drive) and current evolution, followed by EFIT for Grad-Shafranov equilibrium, and then TGYRO to compute the steady-state density, temperature, and (optionally) rotation profiles. These three steps are iterated until a self-consistent, stationary solution is obtained. Since the ohmic current profile will typically diffuse continuously inward causing the on-axis safety factor ($q_0$) to drop, ad-hoc current diffusion can be turned on in the equilibrium step to prevent $q_0$ from falling below one. This crudely mimics the effect of sawteeth on the current-density profile and allows for a stationary solution to be obtained even in inductive plasmas. The flexibility of STEP allows for many variations of this standard workflow.
\begin{itemize}
\item The fixed-boundary solver CHEASE can replace EFIT to provide equilibria for devices that don't have EFIT versions readily available. This is especially important for analyzing future devices where the machine itself is not well-defined.
\item Typically within TGYRO, the neural-network version of EPED is used to give the pedestal structure and the full TGLF and NEO codes solve for the turbulent and neoclassical transport, respectively. STEP can also run the full EPED model as a separate step before TGYRO, for cases where the neural network is not well-trained. Furthermore, a neural-network version of TGLF is also available to accelerate the workflow.
\item A new OMFIT module named CHEF (Current-drive, HEating, and Fueling) is available, which provides increased control over the sources outside of ONETWO by interfacing directly with the FREYA and RABBIT neutral-beam codes, the GENRAY and TORAY RF codes, and a pellet-ablation module (PAM) for core density fueling by pellets. When CHEF is used in this workflow, ONETWO is only used to evolve the current profile.
\end{itemize}

The above workflow and its variations are ``open-loop,'' meaning that a set of sources is predefined and the plasma consistent with those inputs is obtained. We have also written a ``closed-loop'' workflow, wherein the desired $q_0$ and normalized ratio of plasma pressure to magnetic pressure ($\beta_N$) are prescribed and the sources needed to match those parameters are found. To do this, the standard open-loop workflow is wrapped in a root finder with predefined targets and actuators. This has been implemented so that users can choose any set of actuators and targets based on their needs. In this case, $q_0$ and $\beta_N$ are the targets, while an on-axis RF current source and the total neutral-beam power serve as the actuators. The inner, open loop is run to convergence, the solution compared to the target values, and the actuators updated according to the root-finding algorithm. Once a converged solution is found for the outer, closed loop, the user knows both the full plasma state that matches the target parameters, along with the actuators needed to attain that state.

\section{Verification and validation}
\label{sec:v&v}
 
Several studies have been undertaken to demonstrate the accuracy of STEP integrated modeling, which is necessarily limited to the accuracy of the underlying models/codes that are used. These studies include a verification benchmark against other integrated-modeling frameworks and validation against experimental results for a variety of different tokamaks and scenarios.

\subsection{Cross-code benchmark}
\label{sec:benchmark}

\begin{figure}
\centering
\includegraphics[width=0.95\columnwidth]{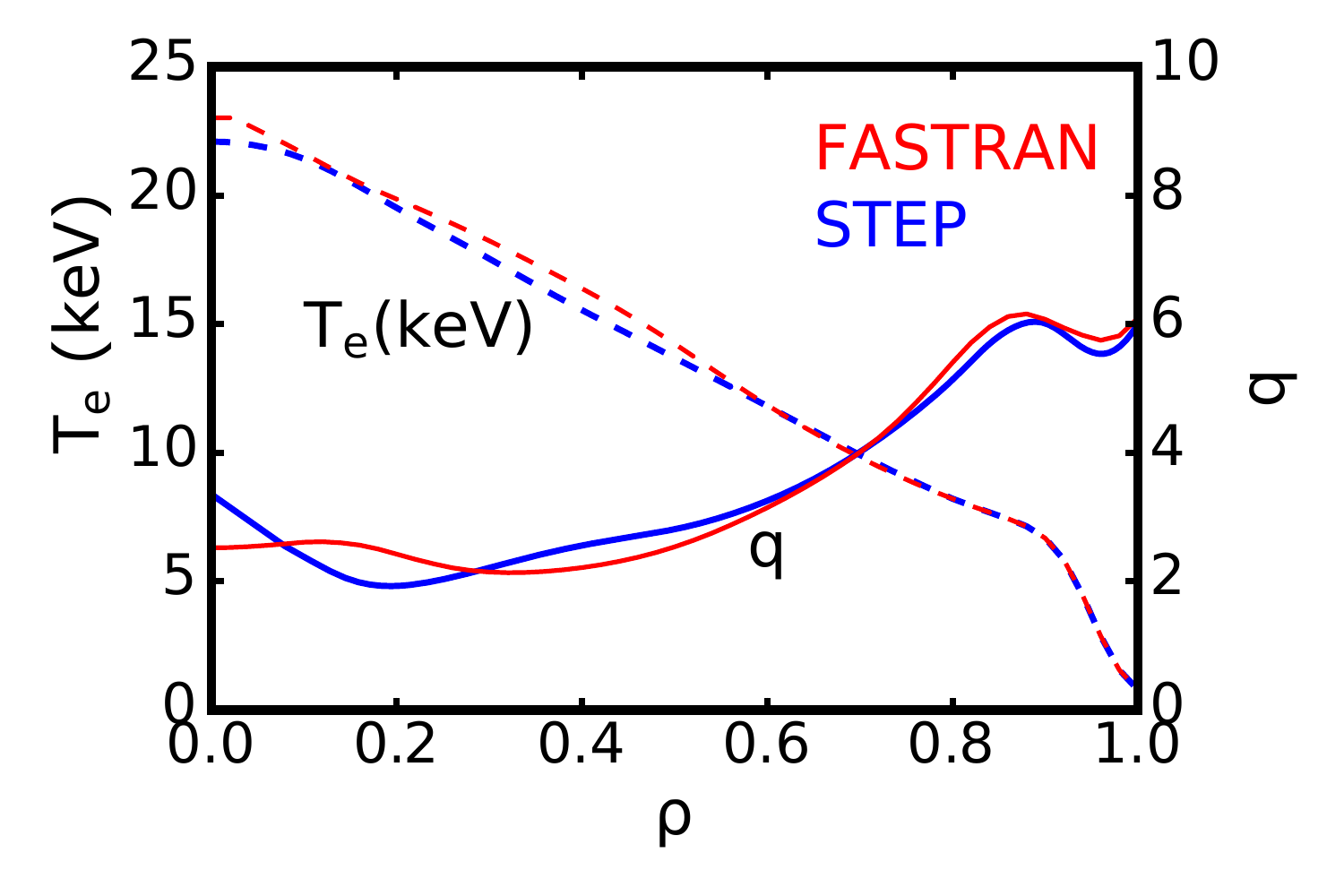}
\caption{Profiles of electron temperature and safety factor for the STEP standard workflow's stationary solution for the ITER weak-shear steady-state benchmark first published in Ref. \onlinecite{murakami:2011}, showing good agreement with the FASTRAN/ONETWO results published in that study.}
\label{fig:benchmark}
\end{figure}

A verification study was previously undertaken by the International Tokamak Physics Activity (ITPA) – Integrated Operation Scenario (IOS) Topical Group, which had six different integrated models simulate an ITER weak-shear, steady-state scenario\cite{murakami:2011}. As can be seen in Figure 1 of Ref. \onlinecite{murakami:2011}, all the integrated models agreed well for key plasma quanties, including densities, temperatures, and safety factor. In order to verify STEP against these results, STEP was initialized using profiles from the FASTRAN\cite{park:2018}/ONETWO simulation of the benchmark. We then ran the standard workflow discussed in Section \ref{sec:workflows}, but using the GLF23 transport model\cite{waltz:1997}, as was used by the other codes in the benchmark, instead of TGLF. In addition, the pedestal was held fixed, at slightly above the peeling-ballooning EPED prediction, as was done by all the codes in the benchmark. Finally, while  FASTRAN/ONETWO used  NUBEAM for NBI and CURRAY for ion-cyclotron RF (ICRF), STEP uses FREYA and GENRAY, respectively. Both  FASTRAN/ONETWO and STEP used TORAY for electron-cyclotron RF. The results for the STEP stationary solution compared to  FASTRAN/ONETWO can be seen in Figure \ref{fig:benchmark}. The difference shown between the two solutions is well within the spread of results found in the original benchmark.

\subsection{Experimental validation}
\label{sec:validation}

\subsubsection{Standard H-mode}
\label{sec:hmode}

\begin{figure*}
\begin{subfigure}{0.49\textwidth}
	\centering
	\includegraphics[width=\textwidth]{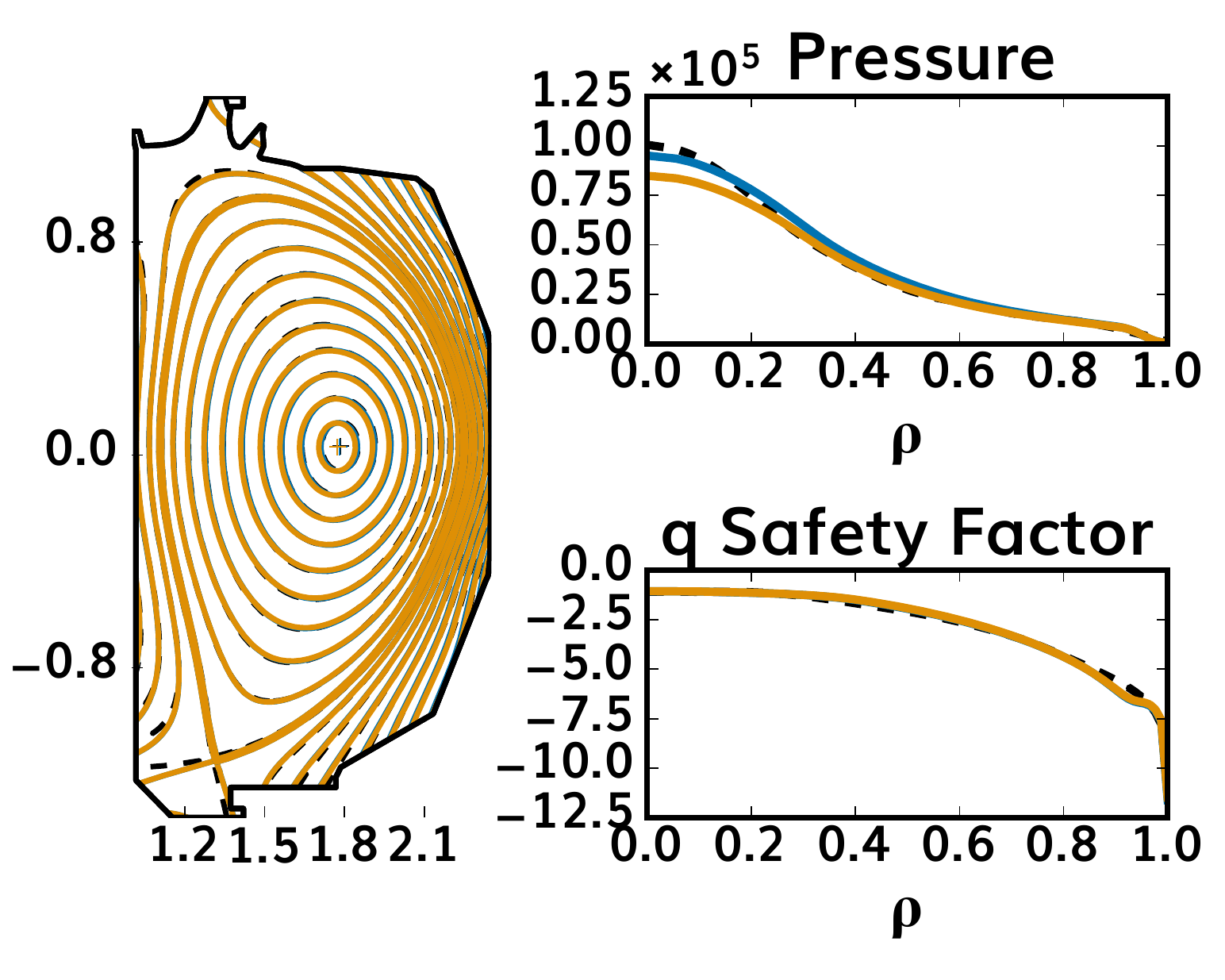}
	\label{fig:175865_equil}
\end{subfigure}
\hfill
\begin{subfigure}{0.49\textwidth}
	\centering
	\includegraphics[width=\textwidth]{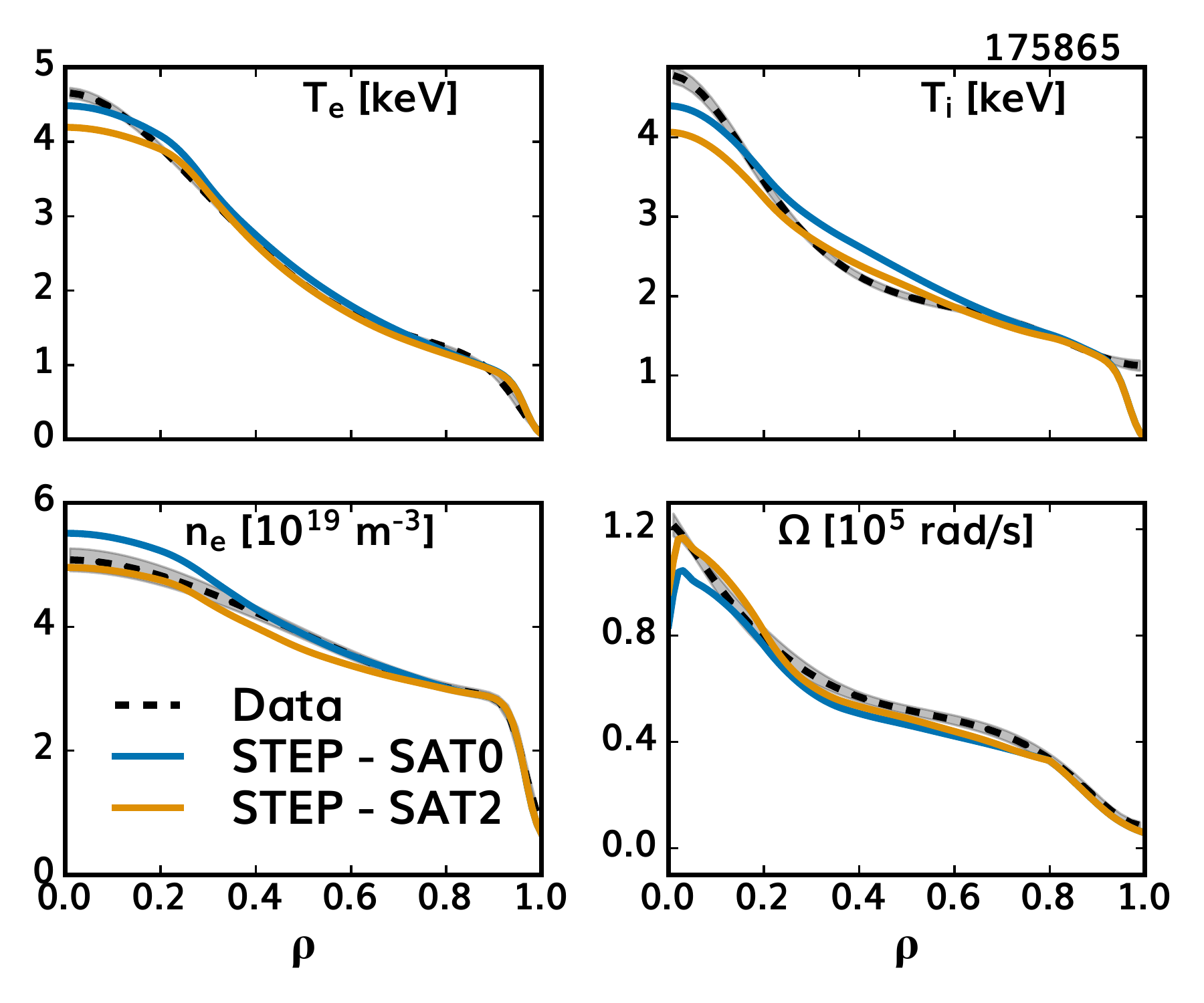}
	\label{fig:175865_profiles}
\end{subfigure}
\caption{STEP accurately reproduces a typical DIII-D H-mode. The dashed black lines show the profile fit to the experimental data, with uncertainties as gray bands where available. Two STEP simulations using the standard workflow are shown in blue (using the SAT0 rule in TGLF) and orange (using SAT2). Good agreement is found in the equilibrium flux surfaces, pressure, and safety factor, as well as the electron temperature ($T_e$), ion temperature ($T_i$), electron density ($n_e$), and toroidal rotation frequency ($\Omega$). For the flux surfaces, the x-axis is major radius and the y-axis is height, in meters. For the other plots, the x-axis is the square-root of the normalized toroidal flux, $\rho$, which is a flux-surface label.}
\label{fig:validate_175865}
\end{figure*}

In addition to showing good agreement with other integrated models, STEP is capable of reproducing experimental plasmas with good accuracy. To validate STEP, we began with experimental data for DIII-D standard inductive H-mode, namely shot 175865 at 2100 ms. This came from a torque-scan experiment in which the NBI power was held constant as the co- versus counter-current NBI was varied. The slice at 2100 ms came from the high-torque (3 N$\cdot$m) phase of the experiment and functions as something close to a typical DIII-D H-mode plasma, with high-rotation and a pedestal limited by type-I edge-localized modes (ELMs). STEP was initialized for an equilibrium reconstruction of this discharge based on kinetic profile data and analysis with  TRANSP\cite{transp:2018, hawryluk:1980, grierson:2018}. The standard workflow described in Section \ref{sec:workflows} was run to stationary convergence with ONETWO for sources and current evolution, EFIT for the equilibrium, and TGYRO with the full TGLF + NEO and EPED-NN. STEP simulations were performed using electrostatic TGLF. Within TGLF, there are three different saturation models that are a progression in accuracy of the fit to the nonlinear gyrokinetic fluctuation intensity. The original SAT0 model\cite{staebler:2005, kinsey:2008} has no coupling between poloidal wavenumbers in its 1D spectrum. The SAT1 model\cite{staebler:2016} is fit to the 2D (radial and poloidal wavenumbers) flux-surface-average intensity and includes mixing of both poloidal wavenumbers and zonal-flow mixing. The SAT2 model\cite{staebler:2020, staebler:2021} is a 3D fit that includes the poloidal-angle dependence of the intensity giving a more accurate fit to the flux-surface shape dependence.  In this validation, we considered two saturation rules, namely SAT0 and SAT2. As can be seen in Figure \ref{fig:validate_175865}, STEP reproduces both the equilibrium and the underlying density, temperature, and rotation profiles with a high degree of accuracy. In particular, SAT2 gives very good agreement with the experimental data, only slightly under predicting the core temperature.

\begin{figure}
\centering
\includegraphics[width=0.95\columnwidth]{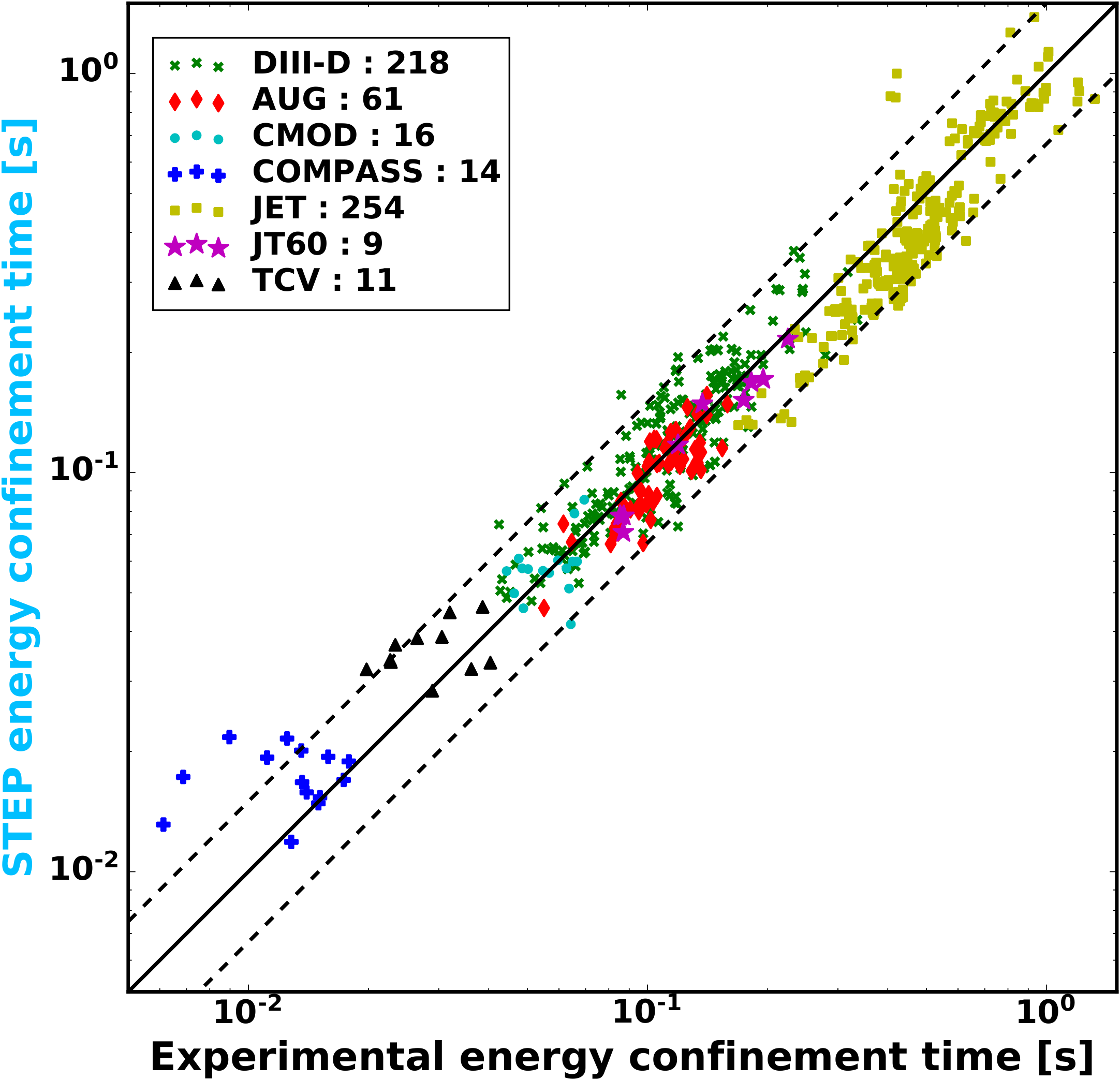}
\caption{The energy confinement time as computed by STEP versus the experimental value for $\sim500$ discharges from seven different tokamaks in the $H_{98,y2}$ database. The dashed diagonal line represents equality, while the solid diagonal line shows the $\pm50\%$ values. Note that the total mean relative error is $\sim19\%$. }
\label{fig:validate_database}
\end{figure}

To demonstrate that this good agreement is not unique to one particular discharge, STEP simulations were run for approximately 500 H-mode discharges from the database used to create the $H_{98,y2}$ energy-confinement-time scaling law\cite{iter:1999}. This database includes data from seven different tokamaks (ASDEX-Upgrade, Alcator C-Mod, COMPASS, DIII-D, JET, JT-60U, and TCV) and spans over three orders of magnitude in confinement time. Like the $H_{98,y2}$ scaling law, these STEP simulations were initialized from only scalar data known for each discharge, namely the plasma current, magnetic field at the geometric axis, line-averaged electron density, effective charge state ($Z_\mathrm{eff}$), major and minor radius, elongation, triangularity, neutral-beam particle energy, and heating power subdivided per heating system. A new Profiles Creator (PRO-create\cite{slendebroek:2023}) module was used to make the initial guess for the plasma state with simple analytic profiles for density, temperature, rotation, and current, along with a self-consistent Grad-Shafranov equilibrium. Each of these plasma states was then analyzed by running the STEP standard workflow with ONETWO, CHEASE, and TGYRO (full TGLF + NEO with EPED-NN) using conservative and identical assumptions on the heating power and other free input parameters. The workflow was iterated until a stationary solution was achieved and the energy confinement time noted for each. Figure \ref{fig:validate_database} plots these STEP-predicted confinement times versus the experimental confinement time, finding a mean relative error of 19\%, less than the 22\% of the $H_{98,y2}$ regression. It's notable that this 19\% error includes obvious outliers at low confinement time from the COMPASS and TCV tokamaks. These discharges were type-III ELMing\cite{zohm:1996}, as opposed to the type-I ELMs assumed by the EPED model, thus explaining STEP's overprediction of their energy confinement times. A more detailed explanation of this work can be found in Ref. \onlinecite{slendebroek:2023}.
%\pagebreak
\subsubsection{Negative central shear}
\label{sec:ncs}

In addition to these standard H-mode cases, STEP is capable of reproducing experimental plasma states in less conventional scenarios. For example, STEP has been used to analyze a negative-central-shear (NCS) plasma from DIII-D. NCS is an advanced-tokamak (AT) scenario under consideration for steady-state reactors. It has both a high bootstrap current fraction and high $\beta$, like most AT scenarios, but increases passive stability by maintaining $q>2$ throughout the whole plasma, eliminating the lowest-order rational surfaces (e.g., 2/1, 3/2), and maintaining large magnetic shear throughout most of the plasma. NCS plasmas have been observed to have internal transport barriers (ITBs) which lead to improved confinement as well. The performance of NCS plasmas is typically limited by macroscopic MHD instability, not transport and available heating power, with the highest achievable $\beta_N$ typically consistent with the theoretical ideal-wall limit\cite{hanson:2017}. We used the closed-loop STEP workflow described in Sec. \ref{sec:workflows} to model an NCS plasma from DIII-D shot 158020. $q_0$ was constrained to a reasonable value as the current diffused, while $\beta_N$ was allowed to float. The inner workflow used CHEF for the sources, ONETWO for current evolution, EFIT for the equilibrium, and TGYRO with the full TGLF + NEO and EPED neural network. As off-axis current was driven in the experiment by ramping the toroidal field throughout the discharge, we implemented this as a non-inductive current source in CHEF based on an analytic derivation\cite{forest:1994}. 

\begin{figure*}
\begin{subfigure}{0.49\textwidth}
	\centering
	\includegraphics[width=\textwidth]{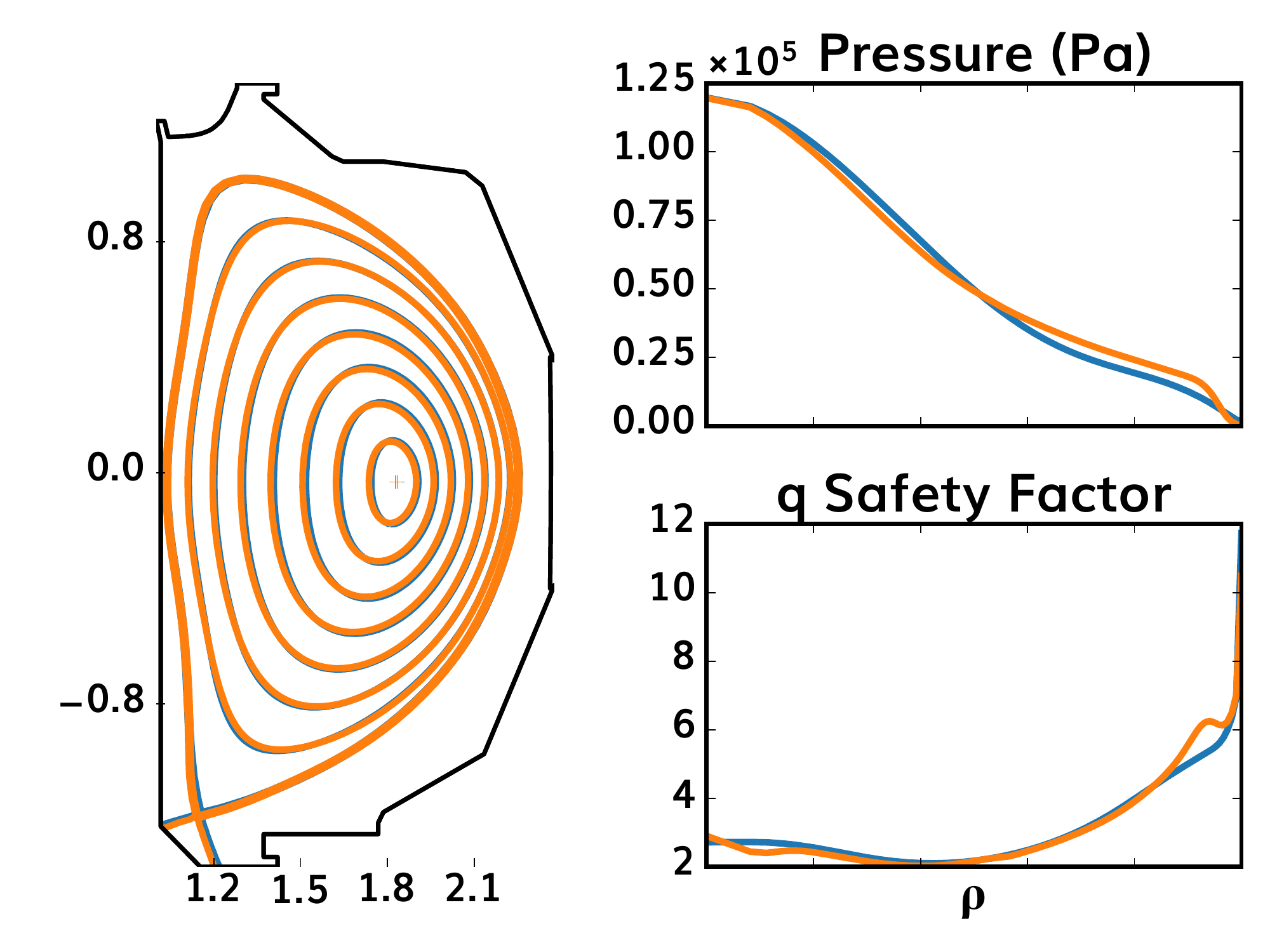}
	\label{fig:NCS_equil}
\end{subfigure}
\hfill
\begin{subfigure}{0.49\textwidth}
	\centering
	\includegraphics[width=\textwidth]{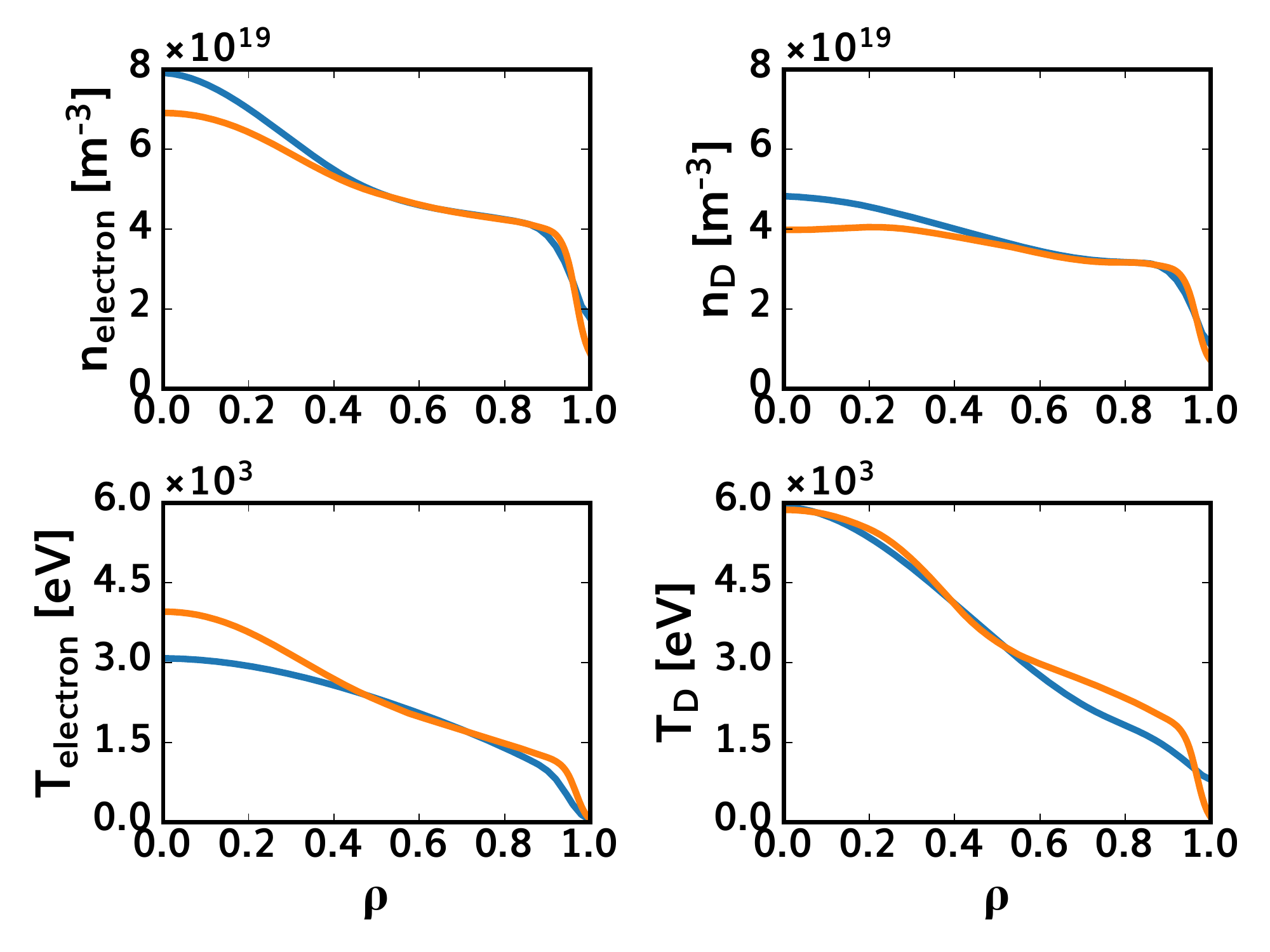}
	\label{fig:NCS_profiles}
\end{subfigure}
\caption{STEP modeling of a DIII-D negative-central-shear plasma in orange, showing good agreement with the experimental reconstruction (blue) for the equilibrium flux surfaces, pressure, and safety factor, as well as the electron density ($n_e$), deuterium (main ion) density ($n_D$), electron temperature ($T_e$), deuterium temperature ($T_D$). For the flux surfaces, the x-axis is major radius and the y-axis is height, in meters. For the other plots, the x-axis is the square-root of the normalized toroidal flux, $\rho$, which is a flux-surface label.}
\label{fig:validate_NCS}
\end{figure*}

TGLF analysis of this discharge was somewhat challenging, necessitating a detailed transport study. We found that SAT0 converges well but significantly overpredicts the core density and electron temperature profiles, finding too strong of an ITB. SAT1 exhibits runaway behavior due to an overestimation of $E \times B$ shear. SAT2 analysis of the experimental reconstruction converges modestly. While iterations of the full STEP workflow then do result in a stationary solution, the TGLF SAT2 analysis of that final state does not appear to be a well-converged transport solution. We found that the best option was to use the SAT1geo saturation rule, which is an intermediate model containing a reduced influence of $E \times B$ shear compared to SAT1 and some of the geometric corrections of SAT2\cite{staebler:2020}. TGLF SAT1geo analysis was able to give converged transport solutions both for the experimental reconstruction and the final STEP stationary state. This STEP solution using TGLF SAT1geo is shown in Figure \ref{fig:validate_NCS}. We find that the predicted equilibrium gives good agreement with the experimental reconstruction, showing a strong ITB at mid-radius with $q>2$ throughout the plasma. The underlying predicted kinetic profiles show a slight overprediction of the electron temperature, compensated in the pressure profile by a slight underprediction of the electron density.

The challenges in the transport analysis of this shot motivate further development of the TGLF saturation rules to predict ITB behavior in NCS plasmas more accurately and robustly.  
%\textcolor{red}{In particular, both SAT1geo and SAT2 appear to underpredict the electron heat flux across the ITB for this plasma and compensate by increasing the electron-temperature gradient to increase trapped-electron-mode (TEM) transport, resulting in higher core electron temperatures than seen in experiment. It has been found that SAT1geo overpredicts TEM transport compared to CGYRO and SAT2\cite{neiser:2023}, and thus requires a smaller increase in ITB height (and better agreement with experiment) compared to SAT2. Thus, the success of SAT1geo for this case is somewhat fortuitous.} 
One promising path will be to use a neural network trained to SAT2 results (SAT2-NN\cite{neiser:2023}) as a preconditioner for running the full SAT2.  Initial tests of this approach have proved promising for improving the convergence of the full SAT2 model.  Nevertheless, the SAT2-NN model is still underdevelopment and any additional transport analysis and future work in this area are beyond the scope of the present study. With respect to STEP integrated modeling, we find that when using the transport model best suited to the analysis of the original experimental discharge,  STEP is able to predict the experimental conditions accurately. This gives confidence that, given an accurate transport model for the desired scenario, STEP has predictive capabilities for AT scenarios.
%\pagebreak
\subsubsection{Negative triangularity}
\label{sec:negd}

\begin{figure}[h!]
%  \centering
  \includegraphics[scale =0.55]{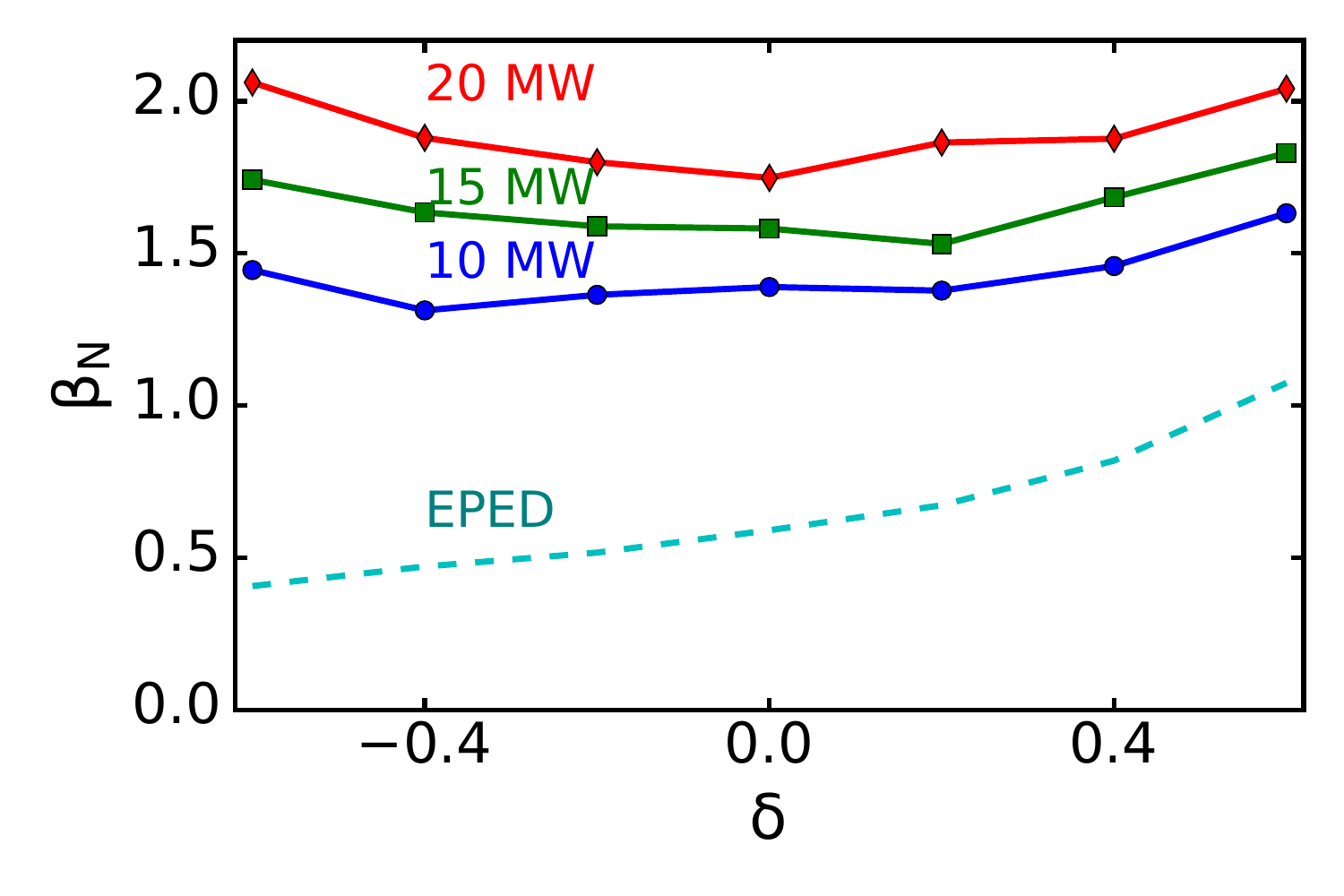}
  \caption{STEP predictions of $\betan$ at different triangularities with input auxiliary power $P_{aux}=10$~MW (blue), $P_{aux}=15$~MW (green), and $P_{aux}=20$~MW (red). The EPED prediction for only the pedestal $\betan$   is shown in teal.}
  \label{fig:negtri}
\end{figure}

Recent experiments on both TCV\cite{camenen:2007} and DIII-D\cite{austin:2019} have found that plasmas with negative triangularity can have H-mode-like confinement with L-mode edges, providing a promising regime for high-performance tokamak operation without edge-localized modes. In order to understand this behavior, STEP modeling has been performed scanning triangularity ($\delta$) using parameters close to DIII-D experiments, namely $\betan=2$, $\neped=0.35\times 10^{20}$ m$^{-3}$, elongation $\kappa=1.6$, toroidal field $B_T=2$ T, and plasma current $I_p= 0.8$ MA. We used a modified version of the standard workflow in Section \ref{sec:workflows}. First, EPED was run for each triangularity assuming $\beta_N=2$. As shown in Figure \ref{fig:negtri}, the EPED-predicted pedestal height increases monotonically with triangularity, from $p_{ped} = 5$ kPa at $\delta=-0.6$ to 12 kPa at $\delta=+0.6$. The significantly weaker H-mode pedestal in \nt{} means that \nt{} will not have a large improvement in confinement from L-mode to H-mode, like \pt{} does. This result is consistent with \nt{} H-mode results on TCV, where a small pedestal and high frequency ELMs were observed when in \nt{} \cite{pochelon:2012}, and consistent with previous peeling-ballooning predictions of a pedestal which showed significant reduction of pedestal height in \nt{} \cite{merle:2017}. Thus, instead of using an L-mode boundary in \nt{} and H-mode in \pt{}, we use EPED boundary conditions for all triangularities, which gives us a reduction in edge confinement as triangularity is reduced without adding the complication of using different edge models.

We then ran STEP with these pedestal boundary conditions for each plasma shape using a workflow of ONETWO, CHEASE, and TGYRO. A radially uniform $Z_\mathrm{eff} = 1.7$ was assumed. TGLF (within TGYRO) was run using the SAT0 saturation rule, the $E\times B$ quench rule, and without electromagnetic effects. The injected auxiliary power used 80 keV NBI with 75\% on-axis co-current and 25\% on-axis counter-current.  A fixed Gaussian source for electron heating was used with 50\% located at $\rho=0.6$ and 50\% located at $\rho=0.0$ to represent electron cyclotron heating (ECH), providing a fair comparison between \pt{} and \nt{} while avoiding the need to optimize gyrotron angles as $\delta$ is adjusted. Figure \ref{fig:negtri} shows the STEP-predicted $\betan$ versus triangularity for three injected powers (10 MW, 15 MW, and 20 MW) assuming a 50/50 mix of NBI and ECH heating.  At $P_{aux}=10$ MW, there is only a weak upturn at \nt{} with $\beta_{N,\delta=-0.6}/\beta_{N,\delta=0.6} = 1.45/1.63 = 0.82$. When  $P_{aux}$ is increased to 15 MW,  the $\beta_N$ at \nt{} becomes closer to \pt{} with $\beta_{N,\delta=-0.6}/\beta_{N,\delta=0.6} = 1.67/1.85 = 0.9$. The STEP prediction of  $\beta_N$  becomes U shaped at $P_{aux}=20$~MW with \pt{} with $\beta_{N,\delta=-0.6}/\beta_{N,\delta=0.6} = 2.01/2.05 = 0.98$. Thus, STEP predicts that confinement improvement at \nt{} will be stronger at high power densities.

\section{Predictive modeling}
\label{sec:reactor}

With confidence in the fidelity of STEP workflows established through the verification and validation presented in Section \ref{sec:v&v}, we now use STEP to make predictions for several proposed or under-construction fusion experiments and reactors.

\subsection{Improvements to 0D reactor studies}
\label{sec:freidberg}

\begin{figure}
\centering
\includegraphics[width=0.95\columnwidth]{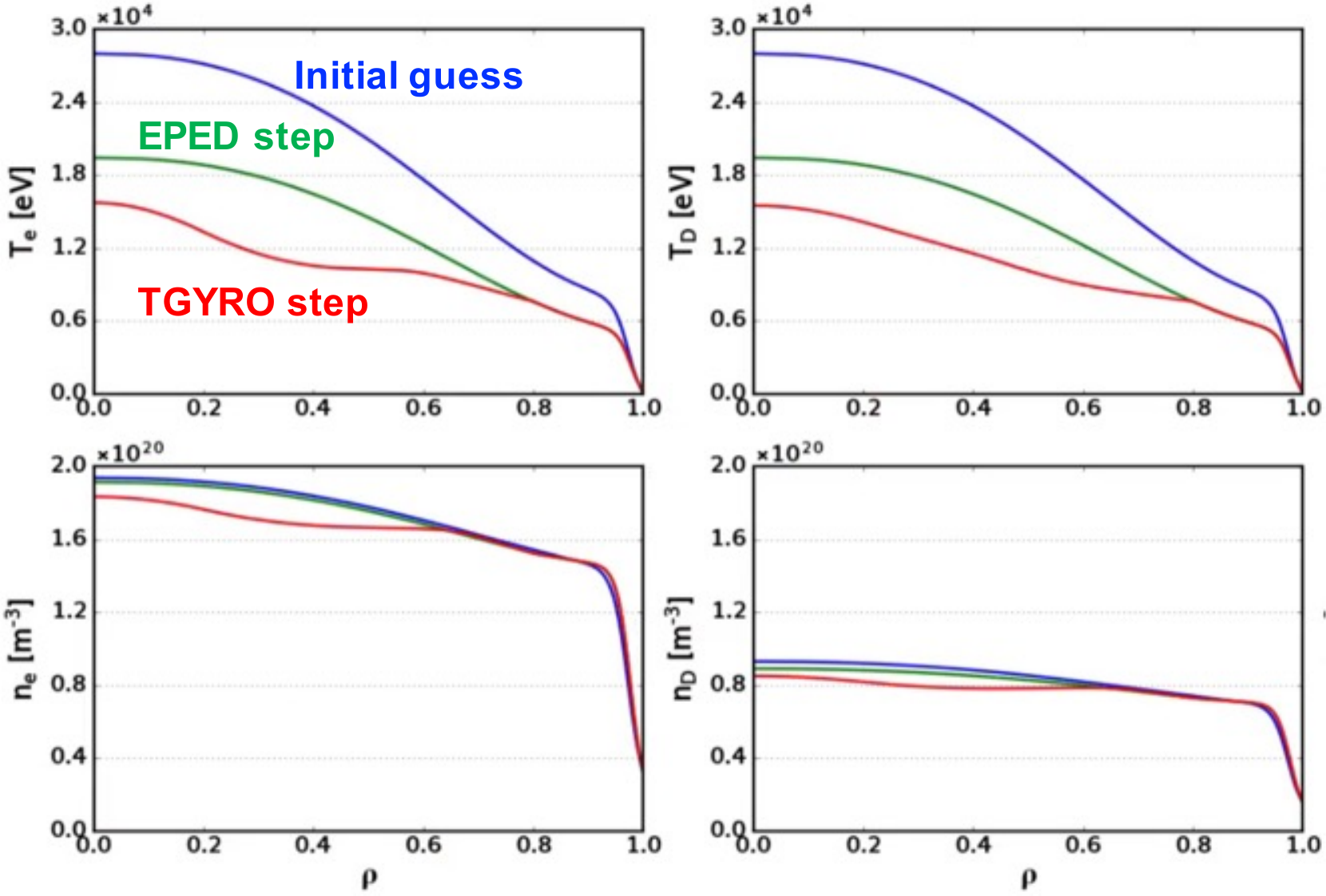}
\caption{Profiles for STEP modeling of a large-aspect-ratio reactor after various steps through one iteration of the standard workflow, for electron temperature ($T_e$), deuterium (main ion) temperature ($T_D$), electron density ($n_e$), and deuterium density ($n_D$). The initial guess is consistent with the 0D scaling used in Ref. \onlinecite{freidberg:2015}. The EPED step predicts a strongly reduced temperature pedestal and then the TGYRO step reduces the core density and temperature further, resulting in significantly reduced fusion power.}
\label{fig:freidberg_profiles}
\end{figure}

While the STEP workflow used for the database validation in Section \ref{sec:validation} can provide a modest predictive improvement over the $H_{98,y2}$ scaling when initialized within the database's existing range of parameters, its real power is to make predictions beyond the database. Zero-dimensional reactor studies often assume confinement that obeys the $H_{98,y2}$ scaling, or some constant factor times the $H_{98,y2}$ value if improved confinement within a scenario is expected. These are only extrapolations; STEP is capable of making physics-based predictions for reactor confinement and performance. To test this, the STEP was initialized using parameters from a 0D reactor study which considered a large-aspect-ratio, ignited tokamak\cite{freidberg:2015}. This study assumed that the confinement time scaled according to the $H_{98,y2}$ regression and found that approximately 1800 MW of fusion power could be achieved in a high-temperature-superconducting reactor with minor radius 0.97 m, aspect ratio 7.65, and on-axis magnetic field 12.4 T. STEP modeling of this case, however, is significantly more pessimistic, finding only 323 MW of fusion power generated. As seen in Figure \ref{fig:freidberg_profiles}, this is primarily caused by a collapse of the pedestal structure. While the initial guess produced by PRO-create based on the 0D parameters in Ref. \onlinecite{freidberg:2015} has a strong temperature pedestal, EPED predicts one significantly lower. Transport analysis with TGYRO then further reduces the core temperature and density. Further iterations of the STEP workflow fail to recover the pedestal height. While the confinement time predicted by STEP is actually $\sim$60\% higher than the $H_{98,y2}$ scaling for the final profiles, the fusion heating power has been reduced significantly and the plasma is no longer in an ignited condition. Additional STEP modeling has found that the fusion power can be increased significantly by reducing the aspect ratio. In fact, even reducing the aspect ratio from 7.65 to 5 (still relatively high) is enough to increase the fusion power to around 1300 MW. Nevertheless, these findings demonstrate the potential limitations of a scaling law approach to reactor design.

\subsection{EXCITE design points}
\label{sec:excite}

\begin{figure}
\centering
\includegraphics[width=0.95\columnwidth]{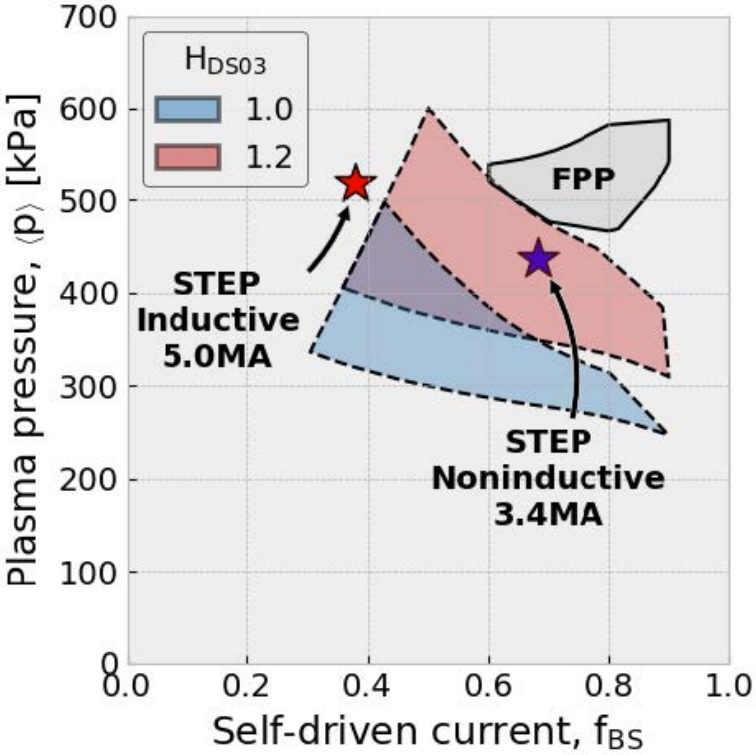}
\caption{Potential solutions for an EXCITE facility in the space of volume averaged pressure and bootstrap-current fraction. The blue, shaded region denotes GASC solutions for EXCITE assuming energy confinement at 100\% of the $H_{DS03}$ scaling while the red, shaded region denotes the same for 120\% of the $H_{DS03}$ scaling. The gray, shaded region denotes expected fusion pilot plant design points. STEP modeling for the an inductive and a noninductive solution are overlaid, showing good confinement and, separately, FPP-like pressure and bootstrap fraction. This figure originally appeared in Ref. \onlinecite{weisberg:2023},  \textcopyright American Nuclear Society, reprinted by permission of Taylor \& Francis Ltd, http://www.tandfonline.com on behalf of American Nuclear Society.}
\label{fig:excite_gasc}
\end{figure}

Recent long-range-planning reports put together by the U.S. fusion research community\cite{cpp:2020}, the Fusion Energy Sciences Advisory Committee\cite{fesac:2020}, and the National Academies of Sciences, Engineering, and Medicine\cite{nasem:2019,nasem:2021} have endorsed a pathway to developing a fusion pilot plant (FPP) in the U.S. in the next two decades. The advanced tokamak, with its significantly improved confinement, is one promising candidate for the FPP. It was noted in several of these reports that there remains a major gap in our present understanding in tokamak physics of how to couple high core performance to effective exhaust solutions (i.e., core-edge integration). A new facility has been proposed called EXCITE (EXhaust and Confinement Integration Tokamak Experiment) that would push plasma-physics parameters near reactor relevancy by going to high pressure and high bootstrap fraction simultaneously. This would allow the U.S. fusion community to test reactor solutions in a non-nuclear environment. Recently, a study was carried out to optimize the design of an EXCITE facility\cite{weisberg:2023}. Initial, zero-dimensional scoping was performed across a wide parameter space using the GA systems code (GASC)\cite{stambaugh:2011}. STEP was then used to analyze two of the proposed 0D solutions in detail, one inductive plasma that would only have high pressure and one noninductive plasma with both high pressure and high bootstrap fraction. The results of the STEP solution are plotted in pressure and bootstrap-fraction space in Figure \ref{fig:excite_gasc}. It is noteworthy that the STEP solution for the noninductive scenario is firmly within anticipated EXCITE operating parameters, while the inductive solution, though at even higher pressure, has somewhat too low bootstrap fraction. In both cases, STEP predicted higher pressures and lower bootstrap fractions than the initial 0D GASC solution. Good energy confinement was found, approximately 110\% of the $H_{DS03}$ scaling\cite{petty:2003} value. More details on this work, including an analysis of the electron-cyclotron and ion-cyclotron heating and current-drive needed to obtain these solutions, can be found in Ref. \onlinecite{weisberg:2023}. 

\subsection{ITER scenarios}
\label{sec:iter}

\begin{figure}
\centering
\includegraphics[width=0.95\columnwidth]{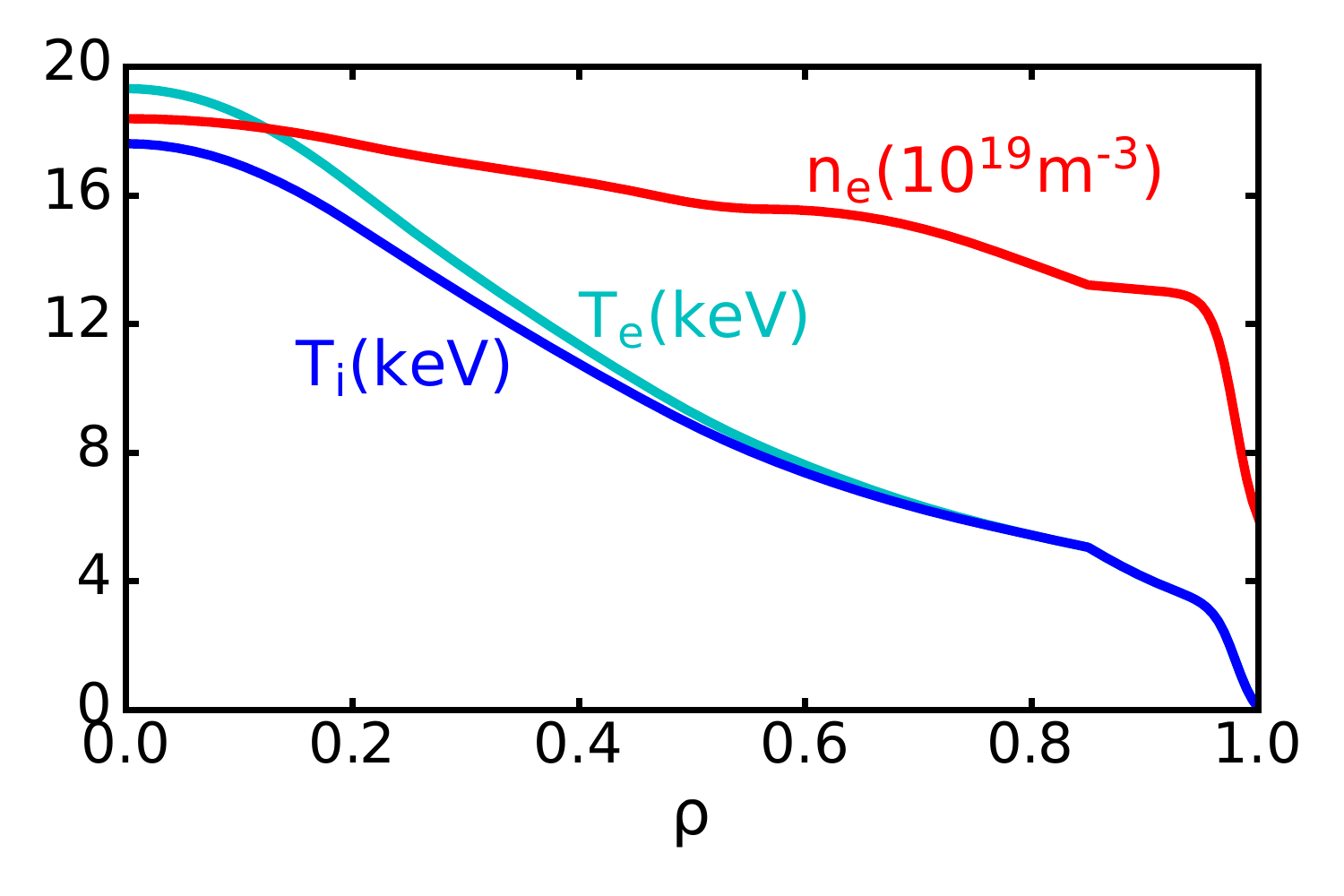}
\caption{Profiles predicted by STEP for the $Q=13$ ITER baseline scenario with a Super-H pedestal and 6 Hz of high-field-side pellet fueling. The electron density ($n_e$) is in red, while the electron temperature ($T_e$) is in solid blue and the ion temperature ($T_i$) in dashed blue. Reproduced from Ref. \onlinecite{knolker:2020}, with the permission of AIP Publishing.}
\label{fig:iter_baseline}
\end{figure}

Making use of CHEF's Pellet Ablation Module (PAM), STEP modeling has been used to study the performance of ITER scenarios with core fueling. Past modeling of the ITER baseline scenario\cite{shimada:2007} found that, by optimizing the pedestal density ($\neped$) and effective charge number ($Z_\mathrm{eff,ped}$), the $Q=10$ milestone could only narrowly be met\cite{solomon:2016, meneghini:2016}, with predicted gains in the range of $10-12$. We performed preliminary STEP modeling of the ITER baseline scenario with 6 Hz pellet fueling from the high-field side, including the effect of the $\nabla B$-drift of the ablated material. This modeling readily found a $Q=13$ Super-H-mode solution at a $Z_\mathrm{eff}=1.8$, the profiles for which are shown in Figure \ref{fig:iter_baseline}. It should be noted that this was done without significant optimization for the pedestal $Z_\mathrm{eff}$, so even higher gains may be possible. More details on this solution are available in Ref. \onlinecite{knolker:2020} and this remains an area for future research.

\begin{figure}
\centering
\includegraphics[width=0.95\columnwidth]{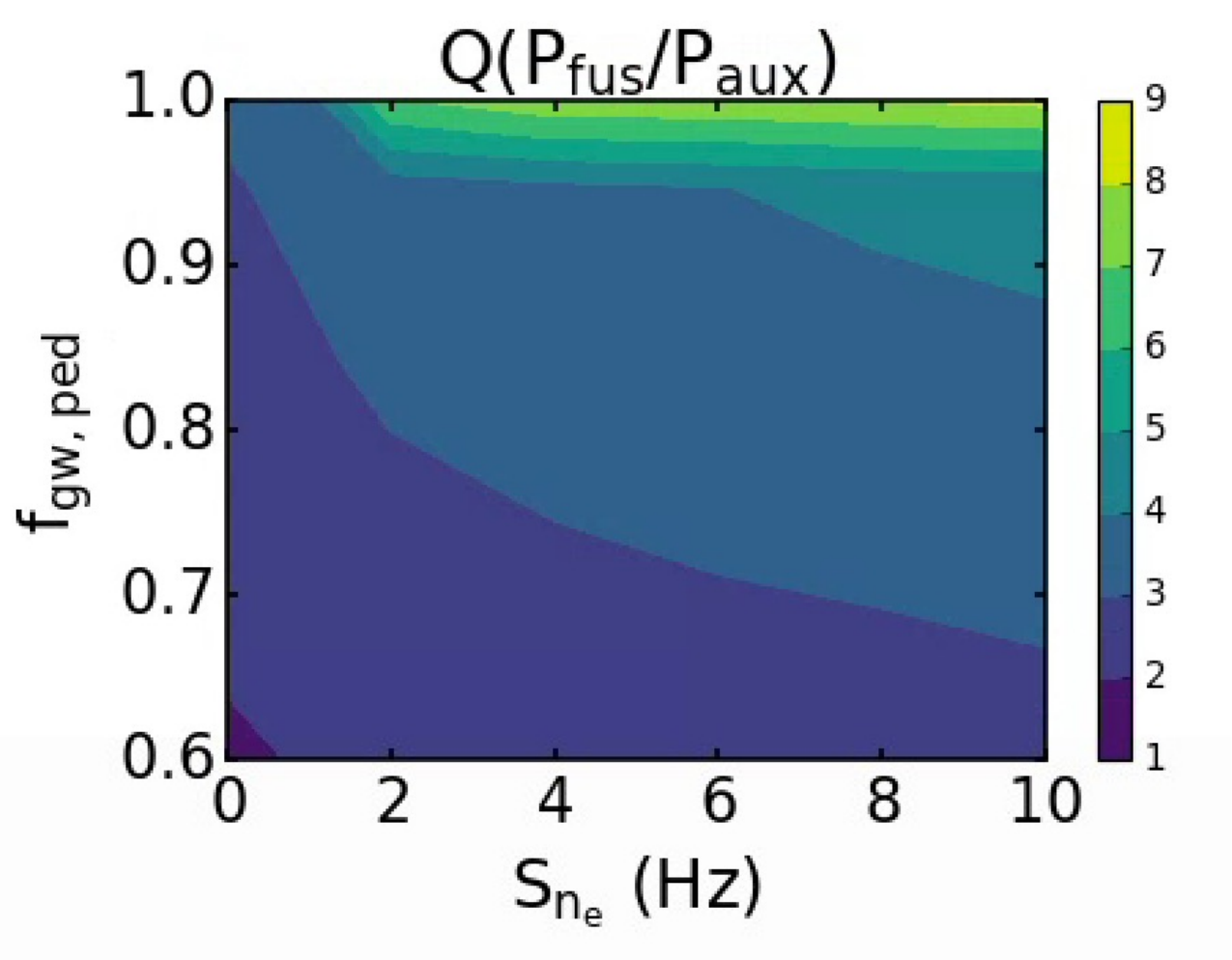}
\caption{Fusion gains as computed by STEP for the 12 MA ITER advanced-inductive scenario, showing a $Q=9$ solution at high Greenwald fraction and high pellet fueling rate.}
\label{fig:iter_advanced}
\end{figure}

In addition to the 15 MA baseline scenario, ITER is also considering advanced-inductive scenarios\cite{luce:2013} which improve performance by pushing stability limits at lower current. STEP modeling was performed for a range of Greenwald-density fractions and pellet-fueling rates for a 12 MA advanced-inductive plasma. As shown in Figure \ref{fig:iter_advanced}, we found that with Greenwald fractions near unity and rapid pellet-fueling rates of 9-10 Hz, high gains of $Q\approx9$ are achievable, approaching the $Q=10$ milestone planned for full-current operation. A more detailed explanation of this work can be found in Ref. \onlinecite{mcclenaghan:2023}.

\subsection{SPARC baseline}
\label{sec:sparc}

\begin{figure*}
\centering
\includegraphics[width=0.95\textwidth]{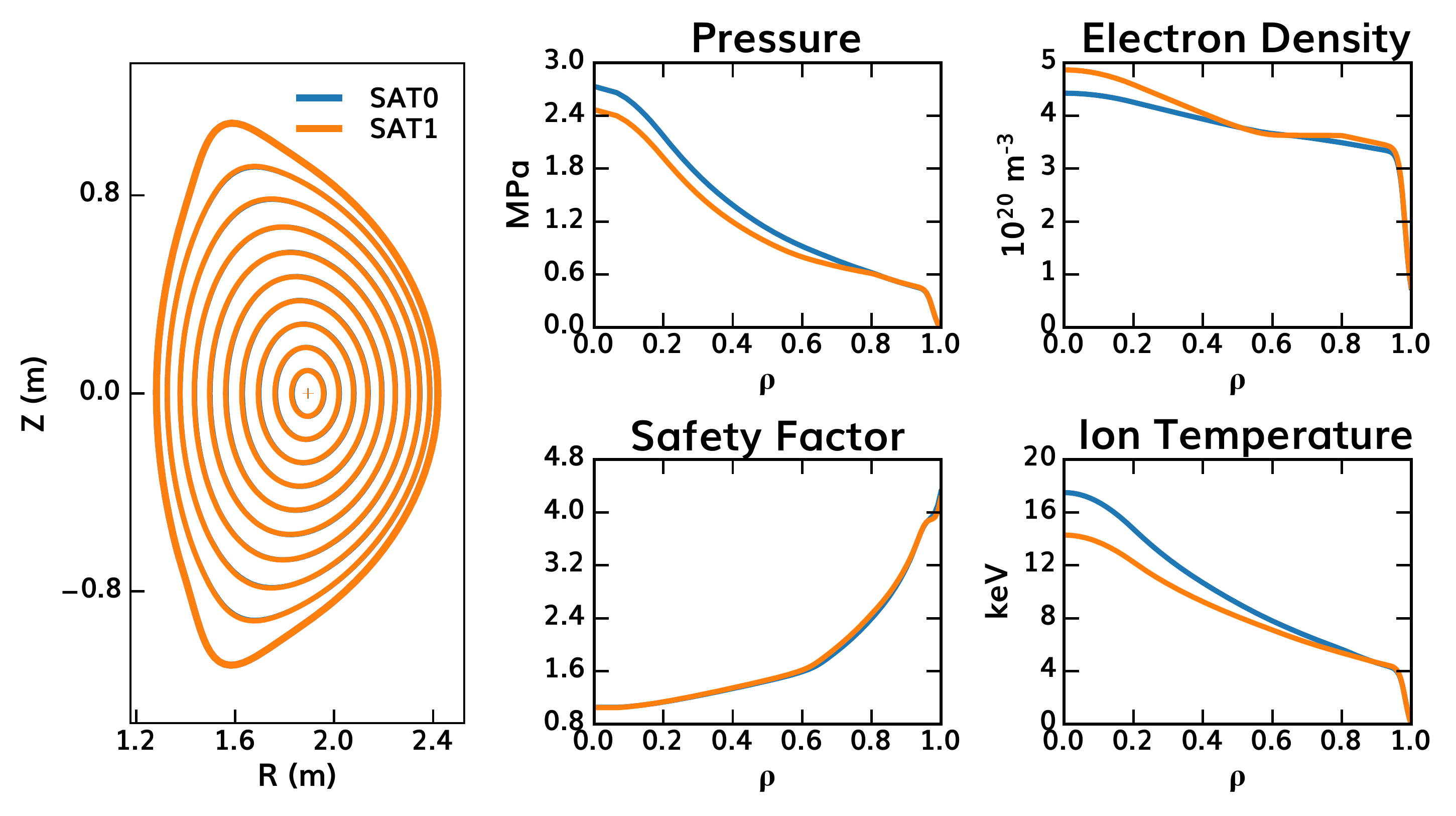}
\caption{Predicted STEP flux surfaces and profiles for the SPARC baseline scenario using the standard workflow with TGLF SAT0 (blue) and SAT1(orange). For the flux surfaces, the x-axis is major radius and the y-axis is height, in meters. For the other plots, the x-axis is the square-root of the normalized toroidal flux, $\rho$, which is a flux-surface label.}
\label{fig:sparc}
\end{figure*}

Finally, a physics basis\cite{greenwald:2020} was recently published for the design of the SPARC tokamak, which is expected to deliver a fusion gain $Q>2$ while demonstrating the use of high-temperature superconducting magnets at reactor scale. A zero-dimensional, empirical scaling analysis with the POPCON code found that the SPARC baseline scenario\cite{creely:2020} would achieve $Q\approx11$, while more detailed, integrated modeling\cite{rodriguez-fernandez:2020} with TRANSP, TGLF, and EPED found that the baseline scenario would have $Q\approx9$. In both cases, this would put SPARC into the burning plasma regime and greatly exceed its $Q>2$ milestone.

We have analyzed the SPARC baseline scenario using the design detailed in the published physics basis. An initial equilibrium was constructed using PRO-create with CHEASE and the SPARC plasma shape\cite{rodriguez-fernandez:2021}. The published SPARC integrated modeling includes deuterium and tritium main ions, $^4$He fusion ash, $^3$He for ion-cyclotron radio-frequency (ICRF) minority heating, tungsten impurity, and a fictitious $Z=9$ impurity to approximate other low-$Z$ ions. Constrained by the need to use real ions in STEP, we initialized the plasma with ion densities as close as possible to what was used in SPARC's TRANSP modeling, namely as a percentage of electron density: $\sim84$\% D-T (50/50 mix), 5\% $^3$He, 1.5\% $^4$He, 0.3\% Ne, and $1.5\times10^{-3}$\% W. This gives a $Z_\mathrm{eff}=1.48$, close the 1.5 in the SPARC baseline.

A standard STEP workflow was used with CHEF for heating and current drive, ONETWO for current evolution, CHEASE for the equilibrium, the full EPED model for pedestal profiles, and TGYRO  with the full TGLF + NEO for core transport. Within CHEF, a GENRAY model of the 120 MHz ICRF antenna concept C\cite{lin:2020} was created with a simplified power spectrum approximating the spectrum found in Ref. \onlinecite{lin:2020}. As SPARC will have two 35-cm-tall antennae just above and below midplane at each toroidal location, we launch several rays spread across a 70-cm region centered on the outboard midplane. 11 MW of power was injected by this antenna and serves as the only auxiliary heating.

The workflow was iterated to convergence using two TGLF saturation rules (SAT0 and SAT1), with ion density fractions reset after each iteration to maintain the initial $Z_\mathrm{eff}$. The final equilibria and profiles for these predictions are shown in Figure \ref{fig:sparc}. We find that with SAT0, STEP predicts $\sim$111 MW of fusion power for $Q\approx10$. With SAT1, STEP predicts $\sim$87 MW of fusion power for $Q\approx8$. This latter value can be compared to the $Q=9.0$ predicted by TRANSP with TGLF SAT1\cite{rodriguez-fernandez:2020}. Given the differences between TRANSP and STEP, this discrepancy seems well within reason. Furthermore, the high gains predicted by STEP give further confidence that SPARC can exceed its $Q>2$ mission with substantial margins.

\section{Conclusion}
\label{sec:concl}

STEP's combination of centralized data exchange and model modularity makes it a flexible, extensible, and accurate tool for modeling steady-state tokamak scenarios. We have shown that STEP is able to reproduce the results of a published integrated-modeling benchmark and can successfully predict the behavior of experimental plasmas, ranging from standard H-modes in seven different tokamaks to negative-triangularity and negative-central-shear plasmas in DIII-D. We also demonstrated STEP's predictive capabilities by simulating several proposed or under-construction fusion devices. Analysis of a large-aspect-ratio reactor showed that 0D analysis can greatly overpredict performance compared to full STEP modeling. STEP simulations of two proposed EXCITE operating points predict higher pressure and lower bootstrap current than a systems code, but find that the noninductive scenario is firmly within desired EXCITE operating parameters.  ITER modeling showed that core fueling by pellets greatly improves performance and could be a key tool for allowing ITER to achieve its Q=10 milestone. Finally, STEP modeling of the SPARC baseline scenario is in good agreement with published TRANSP modeling, showing that the tokamak should be able to greatly exceed its $Q>2$ mission.

While the ideal-MHD codes DCON and GATO are available within STEP, these are primarily used to analyze a converged STEP solution and not iterated within STEP workflows. We intend to make use of closed-loop STEP workflows to target marginal stability by actuating on heating and current-drive sources. Such work will permit the development of predictive stability maps. In addition, the flexibility and modularity of STEP allow for the possibility of  workflows in which parameters could be optimized subject to certain constraints. For example, one could optimize the magnitude and deposition of RF heating and current-drive power while maintaining ideal MHD stability. Such optimization workflows will be an area of future research.

In the near future, we anticipate using STEP to make further predictions for fusion devices, particularly U.S. fusion-pilot-plant (FPP) design points. We intend to support STEP for the community as part of OMFIT, working to develop new workflows and to couple additional physics modules to increase the breadth and depth of STEP capabilities. As is true with OMFIT in general, the broad user-base will allow for collaborative development and widespread use of new capabilities. Furthermore, by coupling to a lightweight current-diffusion solver and developing appropriate workflows, we intend to develop a time-dependent version of STEP that would be capable of simulating a full tokamak discharge, from start-up through flat-top to ramp-down. Such time-dependent simulations will be a key tool in determining the accessibility of the high-performance, steady-state scenarios predicted by the stationary version of STEP. This will be used as a pulse-design simulator for devices like ITER as well as to design discharges for FPP.

\begin{acknowledgments}

This material is based upon work supported by the U.S. Department of Energy, Office of Science, Office of Fusion Energy Sciences, using the DIII-D National Fusion Facility, a DOE Office of Science user facility, under Awards DE-FG02-95ER54309, DE-FC02-04ER54698, and DE-SC0017992. This research used resources of the National Energy Research Scientific Computing Center (NERSC), a U.S. Department of Energy Office of Science User Facility operated under Contract No. DE-AC02-05CH11231. This study is supported by General Atomics corporate funding. Contributions from O. Sauter were supported, in part, by the Swiss National Science Foundation.

\end{acknowledgments}

\vspace{0.2in}
\noindent\textbf{\normalsize DATA AVAILABILITY STATEMENT}
\vspace{0.2in}

The data that support the findings of this study are available from the corresponding author upon reasonable request.

\vspace{0.2in}
\noindent\textbf{\normalsize DISCLAIMER}
\vspace{0.2in}

This report was prepared as an account of work sponsored by an agency of the United States Government. Neither the United States Government nor any agency thereof, nor any of their employees, makes any warranty, express or implied, or assumes any legal liability or responsibility for the accuracy, completeness, or usefulness of any information, apparatus, product, or process disclosed, or represents that its use would not infringe privately owned rights. Reference herein to any specific commercial product, process, or service by trade name, trademark, manufacturer, or otherwise, does not necessarily constitute or imply its endorsement, recommendation, or favoring by the United States Government or any agency thereof. The views and opinions of authors expressed herein do not necessarily state or reflect those of the United States Government or any agency thereof.

\bibliography{Lyons_PoP22}

\end{document}